\documentclass[12pt]{article}
\usepackage{graphicx}
\usepackage{epsfig}
\usepackage{amsmath}
\usepackage{amsfonts}
\usepackage{verbatim}
\usepackage{amsmath}
\usepackage{cite}

\def\pslash{\hbox{/\kern-.5800em$p$}}

\def\gappeq{\mathrel{\rlap {\raise.5ex\hbox{$>$}}
{\lower.5ex\hbox{$\sim$}}}}
\def\lappeq{\mathrel{\rlap{\raise.5ex\hbox{$<$}}
{\lower.5ex\hbox{$\sim$}}}}
\def\unit{\relax{\rm 1\kern-.26em I}}
\numberwithin{equation}{section}
\newcommand{\beq}{\begin{eqnarray}}
\newcommand{\eeq}{\end{eqnarray}}
\newcommand{\dilog}{\ensuremath{\mathrm{Li}_2}}
\newcommand{\nn}{\nonumber}
\def\ltap{\ \raise.3ex\hbox{$<$\kern-.75em\lower1ex\hbox{$\sim$}}\ }
\def\gtap{\ \raise.3ex\hbox{$>$\kern-.75em\lower1ex\hbox{$\sim$}}\ }
\def\CO{{\cal O}}

\def\CM{{\cal M}}
\def\tr{{\rm\ Tr}}
\def\CO{{\cal O}}

\def\CM{{\cal M}}

\def\be{\begin{equation}}
\def\ee{\end{equation}}
\def\bea{\begin{eqnarray}}
\def\eea{\end{eqnarray}}

\newcommand{\gev}{{\rm GeV}}
\newcommand{\tev}{{\rm TeV}}

\newcommand{\ra}{\rightarrow}
\newcommand{\Mmes}{M_{\rm mess}}

\begin{document}
\pagestyle{empty}
\vspace*{5mm}
\noindent \makebox[\textwidth][r]{\small FERMILAB-PUB-08-343-T} \\
\\

\begin{center}

{\Large\bf $\mathbf{R}$--symmetric gauge mediation}

\vspace{1.5cm}
{\sc Santiago De Lope Amigo$^a$}\footnote{E-mail:  slamigo@physics.utoronto.ca}, 
{\sc Andrew E. Blechman$^a$}\footnote{E-mail:  blechman@physics.utoronto.ca}, 
{\sc Patrick J. Fox$^b$}\footnote{E-mail:  pjfox@fnal.gov}, 
{\sc Erich Poppitz$^a$}\footnote{E-mail:  poppitz@physics.utoronto.ca}
\\
\vspace{.9cm}
{\it\small {$^a$Department of Physics, University of Toronto\\
Toronto, ON M5S 1A7, Canada}}\\
\vspace{.5cm}
{\it\small {$^b$Theoretical Physics Department, Fermi National Accelerator Laboratory\\
 Batavia, IL 60510,USA}}\\
\end{center}

\vspace{1cm}
\begin{abstract}
We present a version of Gauge Mediated Supersymmetry Breaking which preserves an R-symmetry---the gauginos are Dirac particles, the A-terms are zero, and there are four Higgs doublets. This offers an alternative way for gauginos to acquire mass in the supersymmetry-breaking models of Intriligator, Seiberg, and Shih.  
We investigate the possibility of using R-symmetric gauge mediation to realize the spectrum and large sfermion mixing of the model of Kribs, Poppitz, and Weiner.  

\end{abstract}

\vfill
\begin{flushleft}
\end{flushleft}
\eject
\pagestyle{empty}
\setcounter{page}{1}
\setcounter{footnote}{0}
\pagestyle{plain}

\section{Introduction}

\subsection{Motivation}
Supersymmetry is one of the most studied ideas for physics at  the LHC. 
Supersymmetric phenomenology is usually described by the minimal supersymmetric standard 
model (MSSM) and its variations (xMSSM's), obtained either by adding extra states, usually gauge singlets, or by focusing on certain regions of parameter space. 

It was only recently
realized \cite{Kribs:2007ac}  that  a new universality class of supersymmetric particle physics models, characterized 
by an extra $R$-symmetry---which can be continuous or discrete ($\supseteq Z_4$), exact or approximate---is not only phenomenologically viable, but also helps to significantly alleviate the supersymmetric flavor problem and has novel  signatures at the TeV scale.  
A model with an exact $R$-symmetry,  called the ``Minimal $R$-symmetric Supersymmetric Standard Model" (MRSSM) was constructed in \cite{Kribs:2007ac}. It was shown, somewhat unexpectedly, that with the imposition of the new symmetry significant flavor violation in the sfermion sector is allowed by the current data,
even for squarks and sleptons with mass of a few hundred GeV, provided the Dirac gauginos are sufficiently heavy, while the flavor-singlet supersymmetric CP-problem is essentially absent.  Stronger bounds on the allowed flavor violation, obtained by including the leading-log QCD corrections, were subsequently given in \cite{Blechman:2008gu}. The Dirac nature of gauginos and higgsinos and the possibility of large sfermion flavor violation in the MRSSM both  present  a departure from usual supersymmetric phenomenology. 

The analysis of the MRSSM in \cite{Kribs:2007ac} was performed in the framework of an effective supersymmetric theory with the most general soft terms respecting the $R$-symmetry. The place of this model in a grander framework, including the breaking and mediation of supersymmetry, was not addressed in detail. The purpose of this paper is to investigate a possible ultraviolet completion of the MRSSM in the framework of gauge-mediated supersymmetry breaking, with the  hope that an ultraviolet completion will help narrow the choice of parameters  of the effective field theory analysis. The focus of this paper on gauge mediation  is motivated by several recent observations.

First of all, phenomenological studies \cite{Kribs:2008hq} of the MRSSM have shown that Dirac charginos are typically the next-to-lightest supersymmetric particles (NLSPs) in the visible sector. This points toward  a possible small scale of supersymmetry breaking, with the resulting light gravitino allowing a decay channel of the light charginos.
 
Secondly, it has been known \cite{Nelson:1993nf} for a while that models with non-generic superpotentials can have both broken supersymmetry and unbroken $R$-symmetry. More recently, Intriligator, Seiberg, and Shih (ISS) \cite{Intriligator:2006dd} observed that metastable supersymmetry-breaking and $R$-preserving vacua in supersymmetric gauge theories are, in a colloquial sense, quite generic.  Majorana gaugino masses require breaking of the $R$-symmetry; instead we explore the possibility that the gauginos are Dirac and the $R$-symmetry is unbroken. Combined with the fact that these vacua can preserve large nonabelian flavor symmetries, it  makes sense to use ISS models to build $R$-symmetric  models of direct mediation of supersymmetry breaking.

\subsection{The MRSSM}

For completeness, here we recall the main features of the MRSSM as an effective softly broken supersymmetric extension of the Standard Model (SM) with an $R$ symmetry. The most important difference from the MSSM are the extended gauge and Higgs sectors and the $R$-charge assignments. 

The quarks and leptons of the SM and their superpartners are described by $R$-charge 1 chiral superfields, while the $R$-charges of the two higgs doublet superfields, $H_u$ and $H_d$ are zero. To allow for $R$-symmetric gaugino masses, SM-adjoint chiral superfields, $\Phi_{1, 2, 3}$, of $R$-charge 0 are introduced.   An additional pair of Higgs doublets, $R_u$ and $R_d$, of $R$-charge 2 are  needed to allow $R$-symmetric $\mu_{u,d}$-terms. The $R$ symmetry forbids the new Higgs fields from  coupling to SM matter through renormalizable operators. While we will refer to $U(1)_R$ as the ``$R$-symmetry," we should stress that for most phenomenological purposes a $Z_4$ subgroup suffices, while a $Z_6$ is sufficient to forbid soft dimension-5 operators violating baryon and/or lepton number such as $QQQL$ and $QQQR_u$.

The MSSM $\mu$-term is forbidden by the $U(1)_R$, and there are new terms in the superpotential allowed by $U(1)_R$ and the SM gauge symmetry:
\be
\label{wr}
\delta W = \mu_u H_u R_u +  \lambda^u_1 H_u \Phi_1 R_u + \lambda^u_2 H_u \Phi_2 R_u + (u \rightarrow d)~.
\ee
The allowed $R$-symmetric soft terms are: soft scalar masses, Dirac gaugino masses (combining the Weyl gauginos of the gauge supermultiplets with the fermion components of the SM adjoint chiral superfields), holomorphic and nonholomorphic masses for the scalar components of $\Phi_{1,2,3}$, and the usual $B_\mu h_u h_d$ term; the MSSM $A$-terms and Majorana gaugino masses are forbidden. As explained in \cite{Kribs:2007ac}, the Dirac  nature of gauginos, the absence of $A$-terms, and the extended Higgs sector---all features following from the $R$-symmetry---can combine to address flavor problems in various regions of tan$\beta$.

\subsection{Outline}

In this paper, we present a model that uses direct gauge mediation and the metastable solution of ISS to generate the MRSSM.  In the next section, we will discuss the relevant details of the ISS model and how it can be used to generate direct gauge mediation with an $R$ symmetry.  We will also introduce notation for computing masses that will be used throughout the paper.  In Section \ref{soft} we will consider how to use the model presented in Section \ref{model} to generate soft terms in the visible sector.  This section is divided into two parts: contributions from the cutoff scale UV physics, and direct contributions from the messenger sector, that we call ``IR contributions".  At this stage we will also discuss a generalization of the model where we identify the important features of the metastable ISS solution and consider how these essential features can be extracted in a general, phenomenologically viable way.  Then, in Section \ref{spectra}, we present some examples of qualitatively different spectra, and discuss constraints such as perturbativity and tuning.  A thourough study of the phenomenology of these models, such as the details of the EWSB sector, collider signals, dark matter, etc., are left for future work.

\section{ISS and $R$-symmetric direct gauge mediation}
\label{model}

Direct gauge mediation postulates that the SM gauge group $G_{\rm SM}$ is part of the global symmetry of the supersymmetry breaking sector, thus relaxing the need to have a separate messenger sector of supersymmetry breaking.  Dynamical models of direct mediation have been considered in the past, see e.g. \cite{Poppitz:1996fw,ArkaniHamed:1997jv}. The ISS models \cite{Intriligator:2006dd}  of metastable supersymmetry breaking
 are attractive setups for constructing  models of gauge mediation, particularly in the $R$-symmetric setup. As we shall see in this paper,  using ISS as an illustrative example of an $R$-symmetric supersymmetry-breaking/mediation sector will teach us some general lessons  on $R$-symmetric mediation; these open the way for the future study of more general   models with different phenomenology.
 
 The ``electric" (high-energy) ISS model is supersymmetric QCD  with gauge group $SU(\hat{N}_c)$ and $N_f$ flavors of quarks $Q$ and $\bar{Q}$, with a tree level superpotential:
 \be
 W_{el.} = \tr \; m\, Q \bar{Q}.
 \label{welectric}
 \ee
  The dual ``magnetic" (low-energy) theory has gauge group  $SU(N_c)$, $N_c= N_f - \hat{N}_c$, $N_f$ flavors of magnetic quarks $q, \bar{q}$,  gauge-singlets $M$, transforming as $(N_f, \bar{N}_f)$ under the flavor group, and superpotential:
 \be
W_{magn.} = \bar{q}\, \CM\, q +  \tr \; m\, \Lambda\, \CM + \ldots
\label{wmagnetic}
\ee
where the dots denote nonperturbatively generated terms (that are not important in the metastable supersymmetry breaking vacuum) and $\Lambda$ is the duality scale.
As ISS show, there exists a metastable supersymmetry-breaking vacuum in this theory, since the equation of motion for $\CM$ following from (\ref{wmagnetic}): $\bar{q}^i \cdot q_j = \Lambda m^i_j$ (the dot denotes summation over the gauge indices) can not be satisfied, for $N_f > N_c$ and a mass matrix of maximal rank $N_f$, due to the rank condition . The flavor symmetry preserved by the mass terms in (\ref{wmagnetic}) is broken in the supersymmetry breaking vacuum, while an $R$-symmetry, under which $\CM$ has $R$-charge $2$ and $q, \bar{q}$ have $R$-charge $0$, remains unbroken. That  the $R$-symmetry is unbroken follows from the Coleman-Weinberg calculation of \cite{Intriligator:2006dd}, which shows what while the dual quarks get expectation values, the trace of $\CM$, which is  a classical flat direction, does not. 

$R$-symmetry breaking is needed to obtain Majorana gaugino masses. Thus, a lot of the model building using ISS and other supersymmetry-breaking models has focused on breaking the $R$ symmetry, either explicitly or spontaneously; for example \cite{Dine:2006xt,Kitano:2006xg,Murayama:2006yf,Csaki:2006wi,Aharony:2006my,Intriligator:2007py,Abel:2007jx,Abel:2007nr,Zur:2008zg,Giveon:2008ne,Haba:2007rj,Amariti:2007qu,Amariti:2006vk}.
As described in the Introduction, in light of the recent observations of \cite{Kribs:2007ac} on the interesting  phenomenological features of supersymmetric  models with unbroken $R$ symmetry, we explore here the contrary possibility. We build $R$-symmetric models of direct gauge mediation, where gauginos are Dirac, and study their phenomenological consequences.

\subsection{The supersymmetry-breaking/mediation sector}
\begin{table}[tb]
\begin{center}
\begin{tabular}{c|c|c|c}
 & $SU(5)_V$ & $U(1)$ & $U(1)_R$  \\
\hline
$M$ & {\rm\bf Adj}+{\rm\bf 1}& 0 & +2  \\
$X$ & {\rm\bf 1} & 0 & +2    \\
$N$ & {\rm\bf 5} & +6 & +2     \\
$\bar{N}$ & ${\rm\bf \bar{5}}$ & --6 & +2     \\
$\varphi$ & ${\rm\bf 5}$ & +1 & 0  \\
$\bar{\varphi}$ & ${\rm\bf \bar{5}}$ & --1 & 0    \\
$\psi$ & {\rm\bf 1} & --5 & 0     \\
$\bar{\psi}$ & {\rm\bf 1} & +5 & 0    \\
$\Phi$ & ${\rm\bf Adj}^\prime$ & 0 & 0   \\
$M'$ & {\rm\bf Adj} & 0 & 0   \\
\end{tabular}
\end{center}
\caption{Charges of superfields of the supersymmetry breaking and mediation sector.  Note that the chiral superfields $\Phi$ are only adjoints under $G_{\rm SM}$ and not the full $SU(5)_V$, denoted by ${\rm\bf Adj}^\prime$.  In addition to the continuous symmetries indicated, we impose a charge-conjugation symmetry ($C$) under which barred and unbarred fields are exchanged, the $G_{\rm SM}  \subset SU(5)_V$ vector superfields change sign, as does $\Phi$; the fields $M,M',X$ are invariant.  }
\label{charges}
\end{table}

To be more concrete, we consider a simple ISS model allowing for direct gauge mediation.\footnote{Since our purpose here is more to emphasize the general features of $R$-symmetric gauge mediation rather than to construct a model with minimal fine-tuning, in most of this paper we consider this simple example where $G_{SM} \subset SU(5)_V$. More general constructions are possible, perhaps even desirable, and will be discussed later in the paper.} For simplicity, we take $N_c = 1$, $N_f = 6$ ($\hat{N}_c = 5$), as done by \cite{Csaki:2006wi}. The ``magnetic" dual theory is then an O'Raifertaigh model. The supersymmetry-breaking vacuum has a reduced vectorlike global symmetry  $SU(6)_V \ra SU(5)_V$ due to the vevs of the dual squark fields $q$ and $\bar{q}$.  We will describe the model in terms of a set of fields with definite quantum numbers under the unbroken  $SU(5)_V$, 
 related to the ones in (\ref{wmagnetic}) as follows:
\beq
\label{fields1}
\CM = \left( \begin{array}{cc} M & N \cr \bar{N} & X \end{array}\right)~, ~ q = \left(\begin{array}{c} \varphi \cr \psi \end{array} \right)~,~ \bar{q} = \left(\begin{array}{c} \bar\varphi \cr \bar\psi \end{array} \right)~.
\eeq

In addition to the fields in (\ref{fields1}), as we will shortly explain, our  model will also require the introduction of  two other fields which  transform as adjoints under $SU(5)_V$ and carry vanishing $R$-charge.  We will call these fields $M^\prime$ and $\Phi$.  In what follows, it will only be necessary for $\Phi$ to be  an adjoint under $G_{\rm SM}$ rather than the full $SU(5)_V$ symmetry ($\Phi$ will be used to give Dirac masses to the gauginos).  This avoids the need for ``bachelor" fields of \cite{Fox:2002bu}; they can be added with minimal trouble, but in the spirit of minimization of the model, we will leave them out.

The charges of the various superfields of the supersymmetry-breaking/mediation sector under the global $SU(5)_V$ symmetry of the ISS model, the $U(1)_R$ symmetry, and a residual $U(1)$ global symmetry (which is spontaneously broken by the dual squark vevs) are given in Table \ref{charges}.

The spontaneous breaking of $SU(6)_V \ra SU(5)_V$ in the ISS model will leave behind a massless Nambu-Goldstone (NG) boson in the messenger sector.  However, the gauging of $G_{\rm SM}$ explicitly breaks the the full $SU(6)_V$ and the NG boson will acquire a mass.  Since the symmetry is broken in this way we consider the more general case where we
 ``tilt" the couplings in the superpotential in eqn.~(\ref{wmagnetic}) so that the $SU(6)_V$ symmetry is explicitly broken, keeping certain ratios of couplings fixed as would be the case for the gauging of $G_{\rm SM}$, e.g. $\kappa$ in $W_{magn}$ below.  Finally, the most general nontrivial tilting of the superpotential that is consistent with the remaining symmetries is
\be
W=W_{\rm magn}+W_1 \label{ISSmodel},
\ee
where:
\be
\label{eq:generalW}
W_{\rm magn}= \lambda\left( \bar\varphi M \varphi + \kappa^\prime\, \bar\psi X \psi + \kappa\, \bar\varphi N \psi + \kappa\, \bar\psi \bar{N} \varphi \right) - f^2( X + \omega\, \tr M ),
\ee
is the (tilted) ISS superpotential from Equation (\ref{wmagnetic}), while 
\be
\label{Myukawas}
W_1= y\left( \bar\varphi \Phi N - \, \bar{N} \Phi \varphi\right)~,
\ee
are additional terms, which explicitly break the global $U(1)$ of Table 1.\footnote{In a complete $SU(6)_V$ description this term can be thought of originating from a term $\bar q  [\hat{\Phi}, \CM] q$ in the magnetic superpotential (\ref{wmagnetic}), with $\hat{\Phi}$ being an extension of $\Phi$ to $SU(6)_V$, similar to the relation between $M$ and $\CM$ in (\ref{fields1}); this allows following Seiberg duality for the construction of the corresponding electric theory, if such a thing is desired.  However, for the purposes of this paper we will simply treat these terms as additional terms allowed by symmetries, making no assumptions as to the origin of the $y$ couplings.}  The couplings in $W_1$  are needed to generate Dirac gaugino mass. 
For now, we simply postulate a $C$-parity, defined in the caption to Table 1, which explains the relative minus sign in (\ref{Myukawas}); we will come back to this point below.  Notice that we can recover the $SU(6)_V$ limit by setting $\kappa=\kappa^\prime=\omega=1$.  By rephasing fields it is possible to take all the parameters in (\ref{eq:generalW}) and (\ref{Myukawas}) to be real, which we do in the following.

\subsection{Scales of supersymmetry breaking and mediation}
\label{scales}

The $F$-term equations at the SUSY breaking metastable minimum of (\ref{eq:generalW}) give:\footnote{Notice that canonically normalizing $\tr~M$ would require a factor of $\sqrt{5}$ be introduced in Equation (\ref{FtrM}), as in \cite{Csaki:2006wi}.  This factor can be reabsorbed into our definition of $\omega$ and not doing so only serves to clutter the notation, so we will not include it here.  Omitting this factor has no effect on the low-energy phenomenology of the model.}
\bea
\langle \bar{\psi}\psi\rangle&\equiv& v^2= {f^2\over \lambda\kappa^\prime}, \label{psibarpsi}\\
\langle F_{\text{Tr}M}\rangle&=&\omega f^2 \label{FtrM}~.
\eea
We also find $\langle\varphi\rangle=\langle\bar{\varphi}\rangle=\langle N\rangle=\langle\bar{N}\rangle=\langle X\rangle=0$, all with masses near $f$.  The other fields are stabilized at higher order in the loop expansion, as we will see below.

At the minimum (\ref{psibarpsi}, \ref{FtrM}) the scalar mass squared terms are:
\be
\label{eqn:scalarmassmatrix}
\begin{pmatrix} \varphi^* & \bar{\varphi} & N^* & \bar{N} \end{pmatrix} \; f^2 \; \begin{pmatrix}
\dfrac{\lambda\kappa^2}{\kappa^\prime} & -\lambda \omega & 0 & 0 \\
-\lambda \omega & \dfrac{\lambda\kappa^2}{\kappa^\prime} & 0 & 0 \\
0 & 0 &\dfrac{\lambda\kappa^2}{\kappa^\prime} & 0\\
0 & 0 & 0 & \dfrac{\lambda\kappa^2}{\kappa^\prime}~
\end{pmatrix}~ \begin{pmatrix} \varphi \\ \bar\varphi^* \\ N \\ \bar{N}^* \end{pmatrix},
\ee
and fermion masses are:
\be
\label{eqn:fermionmassmatrix}
(\varphi \; N) \;f
\begin{pmatrix}
0 &  \sqrt{\dfrac{\lambda \kappa^2}{\kappa^\prime}} e^{i\xi}  \\
\sqrt{\dfrac{\lambda \kappa^2}{\kappa^\prime}}e^{-i\xi} & 0
\end{pmatrix}   \begin{pmatrix}  \bar{\varphi} \\ \bar{N} \end{pmatrix}  ~.
\ee
Notice that all the masses can be scaled to depend on two variables:
\bea
x&\equiv&\lambda\omega~,\nonumber \\ 
z&\equiv&\frac{\omega\kappa'}{\kappa^2},\label{xz}
\eea
and we can define an overall messenger mass scale:
\be
M_{\rm mess}^2\equiv\frac{x}{z}f^2 \label{Mmes2}~,
\ee
which is independent of SUSY breaking.  From the above mass matrices, we see that the $N,\bar{N}$ scalars as well as the two fermion messengers all have mass $M_{\rm mess}$.  The mass eigenstates of the upper $2\times 2$ block of the scalar mass matrix are:
\bea
\phi_+ & = \frac{1}{\sqrt{2}}\left( \varphi +  \bar{\varphi}^* \right)~,\nonumber \\
\phi_- & = \frac{1}{\sqrt{2}}\left( -\varphi +\bar{\varphi}^*\right)~,
\eea 
and have mass squareds:
\be
m_\pm^2 = (1\pm z)M_{\rm mess}^2.
\label{mpm}
\ee
Hence,  to avoid tachyons, we require $z\le 1$.  In fact, $z=1$ is the $SU(6)_V$ limit where there is a massless messenger, as we can see explicitly from (\ref{mpm}).  Further in our analysis, we will take  $z \sim 0.9$.
We note, from (\ref{mpm}), that there is a significant mass hierarchy in the messenger sector  for small breaking of $SU(6)_V$.

The $F$-term conditions (\ref{psibarpsi}) and (\ref{FtrM}) do not fix the vacuum expectation values of $M$ and $\psi, \bar\psi$:
\bea
\psi&=&ve^{(\eta+i\xi)/v}\label{psi}\\
\bar{\psi}&=&ve^{-(\eta+i\xi)/v}\label{psibar}~.
\eea
At one loop, following the calculation of ISS, we find $\langle M\rangle=\langle\eta\rangle=0$ with masses one order of magnitude down from the messenger scale.  This leaves the massless field $\xi$, which is the NG boson of the spontaneously broken $U(1)$ of Table 1.  The Yukawa terms in $W_1$ break the $U(1)$ symmetry, hence this NG boson will get a mass starting at two loops due to the diagram shown in Figure \ref{pNGB}.  In the Appendix, we calculate the diagram and show that it  generates a potential for $\xi$: 
 \be
V_{\rm eff}(\xi)=-\mu^2v^2\cos\left(\frac{2\xi}{v}\right)\label{Veff-xi}~,
\ee
with positive $\mu^2$ given in (\ref{mutwoloop}).
  Thus, 
 the minimum of the potential is at  $
 \langle \xi\rangle = 0$, leading to the conclusion that the $C$-symmetry of the model is not spontaneously broken, and a mass of $\xi$:
\be
m_\xi^2=\left(\frac{\lambda\kappa y}{16\pi^2}\right)^2\Mmes^2 H(z) \label{mxi},
\ee
where $H(z)$ is given in (\ref{hz}).  Here we simply note that $H(1)=2\pi^2/3$ and vanishes in the supersymmetric ($z=0$) limit.

\begin{figure}[h] 
 \centering
    \includegraphics[width=4in]{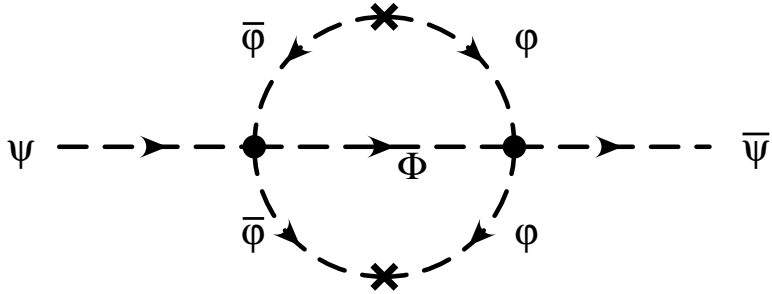} 
    \caption{The leading diagram contributing to the $\xi$ mass.}
    \label{pNGB}
 \end{figure}
 
Finally, we discuss the remaining messenger fermions.  These come with a mass matrix
\be
\label{messengerfermionmassmatrix}
(\psi \; \bar{\psi} \; X) \;v
\begin{pmatrix}
0 &  0 &1  \\
0 & 0 & 1 \\
1 & 1 & 0
\end{pmatrix}   \begin{pmatrix} \psi \\ \bar{\psi} \\ X \end{pmatrix}  ~,
\ee
where $v\sim\Mmes$ is given by (\ref{psibarpsi}) and we have set $\langle\eta\rangle=\langle\xi\rangle=0$.  This matrix can be diagonalized with the result that the $X$ fermion has mass $m_{\tilde{X}}=v$ and the $\psi,\bar{\psi}$ fields mix maximally, one getting a mass $\sqrt{2}v$ and the other with vanishing mass.  This spectrum is not a surprise: 
the $X$ fermion, having $R=+1$, can only mix with the goldstino, the fermionic partner of$\tr M$; one of these fermions marries the gravitino, while the other has a mass $\sim\Mmes$.  The $\psi,\bar{\psi}$ fermions each have $R=-1$ and can mix.  That there is a massless fermion is not surprising, since the $\psi,\bar{\psi}$ sector contains the pseudo-NG boson discussed above, and by supersymmetry this must come with a massless fermion (notice that there is no supersymmetry breaking in these fermion masses).  The $\psi,\bar{\psi}$ superfields can couple to the SM fermions starting at two loops, with gauge fields and messengers in the loops; however, since these operators are generated by gauge bosons the operator is flavor diagonal.  These can then generate four-fermion (flavor conserving) operators that, thanks to supersymmetry, are finite and small, with any divergent loop integrals cut off by the $\xi$ mass (\ref{mxi}).  
Such massless fermions might have some interesting phenomenological or cosmological consequences; from the $R$ symmetry they can only be pair-produced.  We will not say any more about them here.
 
This completes the discussion of the spectrum in the messenger sector.  We may now discuss the phenomenology of the visible sector.  Before doing so we comment on a few technical features of our model.

\subsection{Dirac gaugino masses, $C$-parity, and the extra adjoints } 
\label{cparity}

Generating a Dirac gaugino mass requires a chirality flip on a fermion line as explained in Section \ref{iss-soft}.  This can only come from a superpotential fermion mass term and requires the sum of the $R$-charges of the fields involved to be 2.  The mass of the scalar involved in the loop must be different from the fermion---if not there is a cancellation and the gaugino mass is zero.  This SUSY breaking splitting must come from off-diagonal terms in the scalar mass matrix, otherwise the supertrace is non-zero and there will logarithmic divergences in the scalar masses~\cite{Poppitz:1996xw}.  Only scalar fields of zero $R$-charge can have these off diagonal mass terms.  Hence in order to generate a non-zero Dirac gaugino mass in $R$-symmetric gauge mediation one needs fields with {\it both} $R$-charge 2 and $R$-charge 0.  The model discussed here is of this general form: the $\varphi$ and $\bar\varphi$ have zero $R$-charge and acquire off diagonal masses (\ref{eqn:scalarmassmatrix}), while the fields $N$ and $\bar{N}$ have $R$-charge 2 and supply the needed chirality flip.   We will discuss a more general version of this model, involving fewer adjoints in Section~\ref{sec:generalisedmodel}.

Recall now that in two component notation, Dirac gaugino mass terms are 
\be
m_{1/2} \lambda^{a} \psi^a,
\label{diracmassterm}
\ee
where $\lambda$ is a Weyl  fermion in the adjoint of the gauge group, part of the ${\cal{N}}=1$ vector multiplet, and $\psi$ is the Weyl fermion component of a chiral supermultiplet $\Phi$, also in the adjoint of the gauge group. In addition to preserving an $R$-symmetry, Dirac (\ref{diracmassterm}), as opposed to Majorana, gaugino masses are odd in the gaugino field $\lambda$. Hence, they change sign under $C$-parity if only $\lambda$ is $C$-odd. 
 Note that this already implies that Dirac gaugino masses can not be generated by coupling the adjoint field $M$ from the supersymmetry breaking sector to the gauginos $\lambda$, even in modifications of the ISS model with broken $R$ symmetry, as the field $M$ is even under $C$ (provided the ISS-modification does not break $C$). 
 
 We chose to assign negative $C$-parity to the chiral adjoint $\Phi$ making the Dirac gaugino mass $C$-even. This also requires the relative minus sign between the two couplings in $W_1$ in (\ref{Myukawas}).
Naively, one might think that with a different ratio of the two couplings in (\ref{Myukawas}) the loop-induced Dirac gaugino mass might be reduced or even made to vanish. 
Take, for example, the extreme  case of a positive
  relative sign between the two terms in $W_1$. Then one might argue that the Dirac gaugino mass should vanish: indeed, 
in this case,  we could modify our definition of $C$ so that $\Phi$ was even, thus forbidding  the loop-induced Dirac gaugino  mass term (\ref{diracmassterm}). However, in this case the diagram of Figure \ref{pNGB} would generate 
 a positive-cosine effective potential for $\xi$, instead of (\ref{Veff-xi}), leading to  spontaneous  $C$ symmetry breaking, and giving rise to the same absolute value of the loop-generated Dirac gaugino mass.\footnote{That a maximal absolute value gaugino mass is always generated is true for any value of the couplings in (\ref{Myukawas})---the theory simply wants to maximize the (negative) vacuum energy contribution from gauginos.}

However, a choice of $C$  with even $\Phi$, or the absence of any symmetry, would allow for the generation of a tadpole  for $\Phi_Y$---the gauge singlet hypercharge ``adjoint." Such tadpoles are known to destabilize the hierarchy, see e.g.~\cite{Bagger:1995ay}. Having $\Phi_Y$ odd under an unbroken discrete symmetry eliminates this tadpole, at least the supersymmetry-breaking/messenger sector contribution.  Also, this parity can be used to forbid kinetic mixing of the SUSY-breaking spurion with the hypercharge gauge field strength, which could lead to large tachyonic scalar masses.  
$C$-violation in the SM may introduce other contributions that will involve loops of quarks and leptons and will be suppressed by products of SM gauge and Yukawa couplings.  In what follows we will assume that these contributions are small and can be ignored.  
This is similar to the standard ``messenger parity" that goes along with gauge mediation \cite{Dine:1981gu,Dimopoulos:1996ig,Carpenter:2008wi,Meade:2008wd} except $\Phi$ is also charged under the parity.

The introduction of yet another zero-$R$-charge adjoint, $M^\prime$, of even $C$-parity, is necessitated by the requirement to give the adjoint $M$ a mass. This is because in the absence of $R$ symmetry breaking the fermionic components of $M$ are forbidden from obtaining masses due to loops, as is usually expected in a model where $R$ is broken.

Finally, take note that $G_{\rm SM}\subset SU(5)_V$, and therefore the appearance of these new messengers will have a strong effect on the Standard Model running couplings.  In particular, all the couplings  lose asymptotic freedom and will develop Landau poles.  For typical choices of parameters used below, these Landau poles occur relatively close to the messenger mass scale.

\section{Soft terms in the visible sector}
\label{soft}

Now we proceed to the calculation of the soft terms in the visible sector. To begin, we note that 
there are two main sources of visible sector soft masses in our model:
\begin{enumerate}
\item{ Ultraviolet (UV) contributions due to higher-dimensional operators. 
Typical in models with direct mediation of supersymmetry breaking, all couplings in the SM lose asymptotic freedom. In our model, the scales of the SM Landau poles are not too far above the messenger scale $M_{\rm mess}$. 

As usual, the UV contributions can not be calculated in the low-energy theory. We estimate  the scale suppressing the higher dimensional operators and their contribution  to the  SM soft parameters in Section \ref{uvsoft} using naive dimensional analysis (NDA).  
The largest UV contributions are to soft scalar masses, which are expected to be flavor-nondiagonal, and to $\mu$ and $B_\mu$ terms. UV  contributions to gaugino masses are suppressed, similar to the well-known pre-anomaly mediation gaugino mass problem of supergravity hidden-sector models.}

\item{Infrared contributions to the soft parameters arise due to loops of the particles in the direct-mediation sector and are calculable in the low-energy theory. Messenger loops 
generate Dirac gaugino masses and flavor-diagonal soft scalar masses. The IR contributions to the soft parameters are a loop factor below $\Mmes$  and are calculated in Section \ref{irsoft}. }
\end{enumerate}
There is an interplay between these two types of soft masses in our model. As we discuss in Section \ref{lambda}, the scale suppressing the UV contributions to the soft parameters is about a loop factor above $\Mmes$. Thus the loop-suppressed IR contributions are typically similar to that due to the higher-dimensional operators. This allows us, at the cost of moderate cancellations of the various contributions in the scalar sector (see Section \ref{tuning}) to realize the scenario proposed in \cite{Kribs:2007ac}, where  Dirac gauginos heavier than the scalars  suppress the  flavor-changing neutral currents   
 due to non-degenerate squarks.  

\subsection{Estimating the UV contributions}
\label{uvsoft}

We begin by discussing the typical size of UV contributions.
 From eqn.~(\ref{FtrM}), the $F$-term supersymmetry breaking spurion of $R$-charge 2 is:
 \be 
 \Xi \equiv \langle{\rm Tr} M \rangle = \theta^2 \omega f^2~.
\label{XI}
\ee
 Using this spurion, many $R$-symmetric higher-dimensional operators that communicate supersymmetry breaking to the SM can be written down.  They are all suppressed by some high scale $\Lambda$, the value of which will be discussed later, in Section~\ref{lambda}.  These UV-operator induced soft mass contributions are of order $M_{UV}$, defined as:
 \be 
M_{UV} = { \omega f^2 \over \Lambda} = \left(\frac{z}{\lambda}\right)\left(\frac{M_{\rm mess}}{\Lambda}\right)M_{\rm mess}~,\label{UVmass}
\ee
 where for future use we chose to rewrite $M_{UV}$ in terms of the messenger scale  $M_{\rm mess}^2$ and the dimensionless parameters of eqn.~(\ref{xz})-(\ref{Mmes2}).  

$\Lambda$ is the scale at which these operators are generated and is a model-dependent parameter.  However, before we study the operators that are generated at this scale, a few words can be said about its size.

One possibility is that $\Lambda\sim M_P$: this is the usual expectation from gauge + gravity mediation, where any ``UV operators" are generated by new physics at the Planck scale and are irrelevant.  It solves the flavor problem trivially, since all flavor-changing operators are Planck suppressed; however it assumes that all physics below the Planck scale is flavor-conserving, which is a strong assumption.  As it does nothing to realize the features of the MRSSM, we do not consider this possibility further here.

The other extreme is that $\Lambda$ is related to $\Lambda_3$, the QCD Landau pole, where presumably there is a new dual description that takes over.  It is quite reasonable to assume that there are new states in this dual theory that can generate flavor-violating operators.  We will discuss more careful estimates of $\Lambda$ below but as this turns out to be the most constraining possibility we will consider it throughout the paper.

We now enumerate the  $R$-symmetric  higher-dimensional operators that can be written down.
Dirac gaugino masses $m_{1/2}$ can be generated by the ``supersoft" operator \cite{Fox:2002bu}:
\be
\int d^2 \theta  \; {1 \over \Lambda^3}\;  \tr(W^\alpha \Phi )\; \bar{D}^2 D_\alpha \left(\Xi^\dagger \Xi\right)~ \ra ~ m_{1/2} \sim M_{UV}\left({M_{UV}\over \Lambda}\right)~.
\label{supersoftoperator}
\ee
Similarly, soft scalar masses, $m_{0 \; ij}$, generically flavor non-diagonal, for the SM fields (say, quark superfields $Q$) are given by:
\be
\int d^4 \theta \; {c_{ij}\over \Lambda^2} \; \left(\Xi^\dagger \Xi\right) Q_i^\dagger Q_j ~ \ra ~ m_{0 \; ij} \sim M_{UV}.
\label{softscalaroperator}
\ee
where $c_{ij}$ is a naively flavor-anarchic matrix with $\mathcal{O}(1)$ entries.  
We note that unless $M_{UV}/\Lambda = \CO(1)$,  Dirac gaugino masses due to higher dimensional operators are suppressed\footnote{For $\Lambda \sim M_{Pl}$ this is the well known pre-anomaly-mediation problem of gaugino masses in supergravity without singlets.} compared to the soft scalar mass.  In addition, the smallness of this operator means that we can ignore supersoft contributions to the scalar masses \cite{Fox:2002bu}.  As we will see below, the problem of too-small masses due to higher-dimensional operators will be addressed by direct gauge mediation in this model, along with an estimate of the relevant cutoff scale. 

Next, we recall that in the 
 $R$-symmetric MSSM the usual $\mu$-term is forbidden by $R$-symmetry and that there are, instead,
  two $\mu$-terms, $\mu_u H_u R_u$ and $\mu_d H_d R_d$, where $R_{u,d}$ are two new $R$-charge $2$ Higgs doublets.  
The $\mu_{u,d}$ terms, as opposed to the MSSM, preserve a Peccei-Quinn (PQ) symmetry, which forbids $B_\mu$ but not $\mu_d$, $\mu_u$ ($H_{d,u}$ can be taken to have PQ charge $+2$, $R_{u,d}$ charge $-2$,  and the quark and lepton superfields  charge $-1$). This symmetry implies that, unlike the MSSM,  $\mu_{u/d}$ and $B_\mu$ originate from different operators.

The $B_\mu$ term $B_\mu h_u h_d$ is, however, allowed by $R$ symmetry.  $B_\mu$ 
is generated by an $R$-preserving Giudice-Masiero type operator:
\be
\int d^4 \theta\; {1 \over \Lambda^2}\; (\Xi^\dagger \Xi) \; H_u H_d~\ra ~ \sqrt{B_\mu} \sim  M_{UV}~,
\label{bmuoperator}
\ee
which yields $B_\mu$ similar to the soft scalar mass (\ref{softscalaroperator}). The  $\mu_{u,d}$-terms are instead generated by  $R$-preserving operators\footnote{Notice that C parity implies $\mu_u=\mu_d$, but this need not be required in general; considerations along these lines is delegated for future work.} of the form:
\be
\int d^4 \theta \; {1 \over \Lambda} \; \Xi^\dagger \; H_{u (d)} R_{u (d)} ~ \ra ~ \mu_{u (d)} \sim M_{UV}~.
\label{muoperator}
\ee

In addition, there is an operator that is not forbidden by any symmetry allowing (\ref{bmuoperator}), is renormalizable, naively expected to be unsuppressed and generating an unacceptably large $B_\mu$ term:
\be
\int d^2\theta ~ \Xi H_uH_d~\ra ~ \sqrt{B_\mu} \sim \Mmes~.
\label{badB}
\ee
However, one can put forward arguments in defense of ignoring (\ref{badB}). The only difference between the desirable  (\ref{bmuoperator}) and the undesirable (\ref{badB}) (as written), is that the former vanishes as $\Lambda \rightarrow \infty$ while the latter does not.  Now, the scale $\Lambda$ is expected to be proportional to the SM Landau pole. Thus all UV-suppressed operators coupling the SM to the supersymmetry-breaking sector that we wrote so far---except (\ref{badB})---vanish as one takes the SM gauge couplings to zero, since the Landau pole scale goes to infinity in this limit. One might  adopt   a broad definition of ``gauge mediation" by requiring that all couplings of SM to supersymmetry-breaking sector fields vanish as one takes the SM gauge coupling to zero (and, in our model, the coupling $y$ of $W_1$, which may be related by a high-scale ${\cal N} = 2$ supersymmetry to the gauge coupling). Clearly, imposing this criterion amounts to an assumption on the nature of the unknown UV theory: in particular, it should have an accidental PQ symmetry which forbids (\ref{badB}) but is broken by higher-dimensional operators such as (\ref{bmuoperator}). 
In the absence of an explicit dual, it is hard to be more precise; in practical terms, in what follows we will set the coefficient of (\ref{badB})  to zero and  appeal to technical naturalness in supersymmetry.

The scalars in the adjoint chiral multiplets $\Phi$ (of zero $R$-charge) will also obtain soft masses of order $M_{UV}$ from K${\rm\ddot{a}}$hler potential terms, such as:
\be
\int d^4 \theta \;  {\Xi^\dagger \Xi \over \Lambda^2} \; \left( {\rm Tr} \Phi^\dagger \Phi + {\rm Tr} \Phi^2 \right)~.
\label{Phiscalarmass}
\ee
We can also write a large superpotential ``B term" for $\Phi$ but chose not to for the same reasons as avoiding (\ref{badB}).  
Finally, as explained in Section \ref{cparity}, to avoid  massless SM adjoint fermions, we introduced (see Table 1) another $SU(5)_V$ adjoint, $M^\prime$, of zero $R$-charge. The $R$-preserving operator:
\beq
\label{Mdiracmass}
\int d^4 \theta \; {1 \over \Lambda} \; \Xi^\dagger \tr M M^\prime \ra m_{1/2}^M \sim M_{UV}~,
\eeq
gives rise to  a Dirac mass for $M$ and $M^\prime$ of the same order as the soft scalar mass (\ref{softscalaroperator}).

\subsection{Calculating the IR contributions}
\label{irsoft}

We now consider the calculable IR contributions to the soft mass parameters.  
There is a 1-loop contribution (similar graphs were considered in \cite{Fayet:1978qc}) to the Dirac gaugino mass from graphs involving the $\varphi$ ($\bar{\varphi}$) and $\bar{N}$ ($N$) messengers, shown in Figure~\ref{fig:gauginomass}, as well as the usual two-loop gauge mediated contributions to the scalar masses.  We now proceed to calculate  these soft masses. 

\subsubsection*{Soft masses in ISS-models}
\label{iss-soft}

\begin{figure}[h] 
    \centering
    \includegraphics[width=4in]{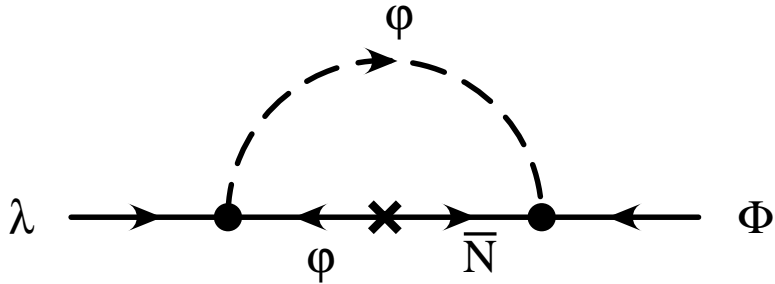} 
    \caption{One of the diagrams contributing to the 1-loop gaugino mass.  The other graphs are obtained by different choices of $\varphi$, $\bar{\varphi}$ $N$, and $\bar{N}$ running in the loop.}
    \label{fig:gauginomass}
 \end{figure}
 
The diagram of Figure~\ref{fig:gauginomass} involves an $R$ preserving fermion mass insertion and a scalar with a SUSY-breaking mass and generates a Dirac gaugino mass. Using the values for our masses and couplings of Section \ref{scales}, we find that the loop-induced Dirac gaugino masses can be written as:
\be
m_{1/2}=\frac{gy}{16\pi^2}M_{\rm mess}R(z)\cos\left({\langle \xi\rangle\over v}\right)~, 
\label{IRgauginomass}
\ee
where:
\be
R(z)=\frac{1}{z}\left[(1+z)\log (1+z)-(1-z)\log (1-z)-2z\right]~,\label{gauginomassfun}
\ee
where $z$ is defined in (\ref{xz}) and measures the off-diagonal supersymmetry breaking mass splitting in the scalar mass matrix (\ref{eqn:scalarmassmatrix}).
Notice the dependence of the gaugino mass on $\cos(\langle \xi\rangle /v)$.  Since  (see discussion in Section \ref{cparity} and the Appendix) $\langle\xi\rangle=0$ this factor is just $1$.  
In principle the $SU(3)$, $SU(2)$ and $U(1)$ pieces of the $SU(5)_V$ may have different $\kappa$ and $y$ coefficients.  However, for simplicity, we take the 3-2-1 pieces to all be the same; breaking this would effect the relative size of the gauginos and sfermions associated with each SM group.

The sfermions acquire a gauge-mediated mass from loops involving the messengers $\varphi,\, \bar{\varphi}$, but not $N,\,\bar{N}$ since they do not have supersymmetry-breaking masses.  Following \cite{Martin:1996zb}, this contribution can be calculated.  Thus, the contribution from gauge group $a$ to a sfermion mass squared is:
\be
m_0^{(IR)2}=2C^{(a)}_F\left(\frac{\alpha_a}{4\pi}\right)^2M_{\rm mess}^2F(z)~,
\label{IRscalarmass}
\ee
where $C^{(a)}_F = (N^2-1)/2N$ for $SU(N)$ and $\frac{3}{5}Y^2$ for $U(1)_Y$ and:
\be
F(z)=(1+z)\left[\log(1+z)-2\dilog \left(\frac{z}{1+z}\right)+\frac{1}{2}\dilog\left(\frac{2z}{1+z}\right)\right]+(z\rightarrow -z)~.
\ee

We note that the contribution of the $R$-symmetric messenger sector to soft scalar masses (\ref{IRscalarmass})  is the same as that of one messenger multiplet in usual gauge mediation. The function $F(z)$ from (\ref{IRscalarmass}), with our parameter $z$ identified with   $F/\lambda S^2$ of usual gauge mediation, is the same appearing in, e.g., \cite{Martin:1996zb}. The  Dirac gaugino mass (\ref{IRgauginomass}), however, is governed by a different function of $z$ than in the case of Majorana mass. This qualitative difference arises because the Dirac gaugino mass requires the presence of an $R$-preserving chirality flip in the loop. This $R$-symmetric chirality flip does not appear in the two-loop diagrams generating the scalar mass, which are thus identical to those in one-flavor gauge mediation---the messenger scalars $\varphi$ and $\bar\varphi$, which have a supersymmetry-breaking spectrum contribute to the scalar masses, while $N$ and $\bar{N}$, which are supersymmetric, do not.  

\begin{figure}
\centering
    \includegraphics[width=4in]{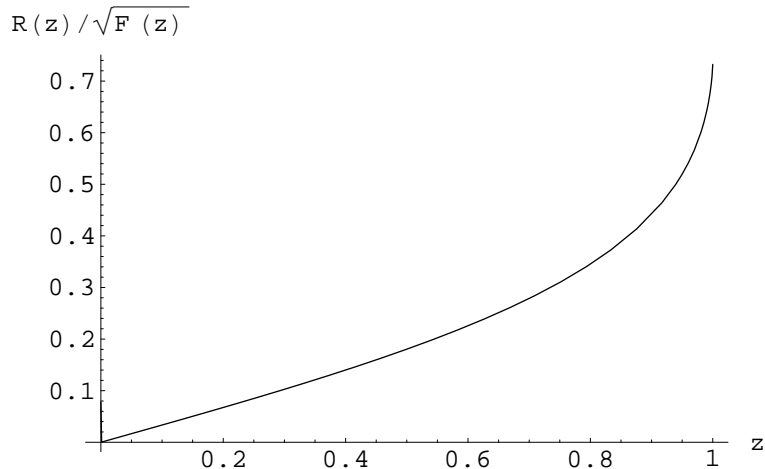} 
    \caption{The function of $z$ entering the ratio (\ref{IRgauginoscalarratio}) of gaugino to scalar mass.  At $z=1$ there is a branch point, with $\left|R(z)/\sqrt{F(z)}\right|\rightarrow0.732$ as $z\rightarrow 1-$.}
    \label{massratio}
 \end{figure}

In addition note that $|R(z\rightarrow 0)|\rightarrow z^2$, unlike usual gauge mediation where $m_{1/2}\sim z$.  This is easy to understand since due to R-charges the Dirac mass operator  (\ref{supersoftoperator}) needs two insertions of the spurion, $\omega\,f^2$, unlike a Majorana mass that needs just one insertion.  This qualitative difference leads to the general fact that in $R$-symmetric gauge mediation the gaugino mass is typically lighter than the scalar mass, in contrast to usual gauge mediation, where the $m_{1/2}:m_0$ ratio is larger than unity, see \cite{Martin:1996zb}.  The ratio of gaugino to sfermion mass in $R$-symmetric gauge mediation is: 
\be
\frac{m_{1/2}}{m_0^{IR}} =\frac{1}{\sqrt{2C_F}}\left( \frac{y}{g}\right)\left(\frac{R(z)}{\sqrt{F(z)}}\right)~.
\label{IRgauginoscalarratio}
\ee
The ratio $R/\sqrt{F}$, as Figure \ref{massratio} shows, is strictly less than $1$: for $z=0.99$, $|R/\sqrt{F}| =.64$.  Thus, in order to solve the supersymmetry flavor puzzle along the lines of \cite{Kribs:2007ac}, which requires large gaugino to squark mass ratios, within an ISS supersymmetry-breaking-cum-mediation sector, we must have a large  Yukawa coupling $y$ (near the boundary allowed by perturbativity, as we will discuss in Section \ref{tuning}).

Finally, the scalar adjoint fields in the $\Phi$ supermultiplets also get real and holomorphic masses from the messenger loops:\footnote{We thank Markus Luty for pointing this out to us.}  
\[V_{\rm eff}=m_\phi^2\phi^*\phi+\frac{1}{2}B_\phi(\phi^2+\phi^{*2})~.\]
These are given by 
\bea
m_\phi^2&=&\frac{y^2}{16\pi^2}\Mmes^2 R_s(z)~, \label{madj}\\
B_\phi&=&\frac{y^2}{16\pi^2}\Mmes^2 R(z)~, \label{Badj}
\eea
where
\be
R_s(z)=\frac{1}{z}\left[(1+z)^2\log(1+z)-(1-z)^2\log(1-z)-2z\right]~,
\ee
and the $z$ dependence in (\ref{Badj}) is the same as in the gaugino mass (\ref{gauginomassfun}).  These masses are the same order, but it can be seen that $|B_\phi|<m_\phi^2$ for any value of $z$, so the gauge symmetry is protected.  Also notice that $B_\phi$ is strictly negative, which means that the scalar will always be lighter than the pseudoscalar.  This is the reverse of ordinary supersoft mediation \cite{Fox:2002bu}.  Notice that since this a one-loop scalar mass, it is enhanced compared to the gaugino mass:
\be
\frac{m_\phi}{m_{1/2}}\sim\sqrt{\frac{4\pi}{\alpha}}~,
\ee
where $\alpha$ is the fine structure constant of the relevant gauge group.  Thus we generally expect the adjoint scalars to be roughly an order of magnitude heavier than the gauginos, although there could be a sizable cancellation between the real and holomorphic mass.  In addition, there could be cancellations with the UV operators that we defined in the previous section (\ref{Phiscalarmass}).

\subsubsection*{Generalized $R$-symmetric gauge mediation}
\label{sec:generalisedmodel}
 
In this section, we introduce a model of generalized $R$-symmetric gauge mediation. Inspired by previous discussions of generalized gauge mediation, see e.g. \cite{Martin:1996zb}, we implement supersymmetry breaking in terms of an $R$-charge 2 spurion
$\Xi \sim \theta^2 f^2$, instead of a dynamical supersymmetry-breaking sector. 

From the ISS model considered in the previous sections, we learned that only the fields $\varphi, \bar{\varphi}$, and $N, \bar{N}$ of  ISS  play a role in the mediation of supersymmetry breaking to leading order in the loop expansion.  Furthermore,  as we  explained in Section \ref{cparity}, this is the minimal set of messenger fields required  to achieve $R$-symmetric gauge mediation. Thus, in our generalized model, we will keep only these fields and consider a messenger sector consisting of $N_{\rm mess}$ copies: 
\be
W_{mess}= \sum\limits_{i = 1}^{N_{\rm mess}} \left( \Xi \; \bar\varphi^i  \varphi^i + M_{\rm mess} \;\bar\varphi^i N^i  + M_{\rm mess} \; \bar{N}^i \varphi^i 
+  y \; \bar\varphi^i \Phi N^i -y \; \bar{N}^i \Phi \varphi^i\right)~.
\label{Wtoy}
\ee
Here $\Xi$ is the supersymmetry breaking spurion (\ref{XI}), $M_{\rm mess}$ is a rigid messenger mass scale; the $R$-assignments of the multiple copies of messengers are the same as their namesakes of Table 1, as is their $C$-parity. 

The messenger sector (\ref{Wtoy}) gives rise, through the same set of one and two loop diagrams as the ones discussed in the previous section, to $N_{\rm mess}  \times$  the gaugino mass contribution of (\ref{IRgauginomass}) and $N_{\rm mess} \times$ the scalar mass squared contribution of (\ref{IRscalarmass}), where we reinterpret $z=f/\Mmes$. Thus the ratio of loop-induced gaugino to scalar mass of eqn.~(\ref{IRgauginoscalarratio}) 
is enhanced by a factor of $\sqrt{N_{\rm mess}}$.
 The enhancement of the Dirac gaugino mass by $\sqrt{N_{\rm mess}}$ in the generalized model  relaxes (some of, see Section \ref{tuning}) the need of  having a large Yukawa coupling $y$. In addition, the absence of the SM adjoints $M$, $M^\prime$ from  (\ref{Wtoy}) pushes the SM  Landau pole up: we note that  the $\alpha_s$ beta function  of the MRSSM  vanishes already above the scale of the Dirac gaugino mass and thus adding any colored messenger inevitably leads to a Landau pole. We will have to say more about this below. 
 
To end this section, we note that in light of its phenomenogical desirability, it would be of some interest to have a UV completion of the  generalized messenger model of (\ref{Wtoy}), ideally  including both the dynamics of supersymmetry breaking and generating the messenger mass scale $M_{\rm mess}$ without introducing the extra adjoint baggage of the ISS model; we leave this for future work.\footnote{One simple way to achieve this is to make $\Xi$ dynamical and add a linear term  $f^2 \Xi$ to (\ref{Wtoy}). The model with superpotential $W_{mess} + f^2 \Xi$ has an $R$-preserving supersymmetry-breaking   (possibly metastable) vacuum  at the origin of moduli space ($\Xi=0$) as a consequence of the $R$-charge assignments \cite{Shih:2007av}.  One could further   ``retrofit" \cite{Dine:2006gm} the explicit mass scales.}

\section{Numerics}
\label{spectra}

\subsection{How high can $\Lambda$ be?}
\label{lambda}

It is well known that in order to avoid constraints from $K-\bar{K}$ mixing, the dimension six operator
\[\int d^4\theta\frac{Q^\dag QQ^\dag Q}{\Lambda_Q^2}\]
must have a cutoff $\Lambda_Q\sim 10^3$ TeV.   Thus we need to chose parameters such that our cutoff is no smaller than this.
 
To understand how large the scale suppressing the UV contributions ($\Lambda$) can be, we must consider the location of the Landau pole.  Consider the beta functions of $G_{\rm SM}$:
\beq
{d \over d \ln \mu} \; {1 \over \alpha_i (\mu) } =-{ b_i \over 2 \pi}~.
\label{SMbeta}
\eeq
The model (\ref{ISSmodel}) contains fields that transform under $SU(3)_C\times SU(2)_L$ as:
\bea
M,M^\prime&:&({\bf 8},{\bf 1})+({\bf 1},{\bf 3})+({\bf 1},{\bf 1})+({\bf 3},{\bf 2})+({\bf \bar{3}},{\bf 2})\nonumber \\ 
\Phi&:&({\bf 8},{\bf 1})+({\bf 1},{\bf 3})+({\bf 1},{\bf 1})\nonumber \\ 
\varphi,N&:&({\bf 3},{\bf 1})+({\bf 1},{\bf 2})\nonumber \\ 
\bar{\varphi},\bar{N}&:&({\bf \bar{3}},{\bf 1})+({\bf 1},{\bf 2})~.
\label{mspectrum}
\eea
Now we must consider how the spectrum behaves, since the running will be sensitive to the fermionic and bosonic mass thresholds of the various multiplets.  We solve the one-loop renormalization group equations including the various contributions as we pass their mass threshold, but we do not include finite threshold effects.  

The presence of a large number of fields charged under the SM means that the Landau pole of $SU(3)$ typically occurs at a relatively low scale, resulting in potentially sizeable UV-induced soft masses (\ref{UVmass}).  However, the Dirac gaugino mass will still be too small if the UV-generated operator (\ref{supersoftoperator}) is the only source of its mass.  For the Yukawa couplings in (\ref{Myukawas}) of order one the gauginos have phenomenologically viable masses but the gluino will still be somewhat lighter than the squarks, see (\ref{IRgauginoscalarratio}).  Without a large value for $y$ it is not possible to realize the scenario of \cite{Kribs:2007ac}.  For larger values of $y$, sufficient to allow for large squark mixing and the interesting flavor physics of the MRSSM, there will be a Landau pole for some Yukawas below the strong coupling scale of $SU(3)$.  The generalized model of Section~\ref{sec:generalisedmodel} alleviates some of these issues by removing some of the adjoints,
  raising the Landau pole, and increasing the number of messenger families, which lowers the Landau pole but also raises the gaugino:scalar mass ratio.  

Once the location of the $SU(3)$ Landau pole $\Lambda_3$ has been determined we may estimate the size of the UV contributions.  If all gauge and Yukawa couplings became strong at the same scale one would expect  that the scale $\Lambda$ of Section~\ref{uvsoft} would be related to the strong coupling scale $\Lambda_3$ by, $\Lambda_3\sim 4\pi \Lambda$.  However, not all couplings become strong at the same scale and the operators involve $\Xi$, which is not charged under $SU(3)$.  Such operators should have a suppression from the perturbative coupling which is weak at that scale, weakening some of the constraints that we will find below.

Of course, we know that while there should be a suppression, it is hard to estimate: above $\Lambda_{3}$, in the absence of an explicit  dual description, we have no idea how the other couplings run (as we have a duality cascade, where after dualizing $SU(3)$, the other gauge content will change) and where the other Landau poles now are.  For the purposes of estimating the UV contributions we will therefore make the simplifying but conservative assumption that $\Lambda=\Lambda_3/4\pi$, which potentially overestimates the size of the UV contributions, especially in the electroweak sector.

\subsection{Sample Spectra}

In this section we will consider three examples of spectra: the full ISS model with perturbative Yukawas, the full ISS model with large $y$ (and consequently large gaugino masses) and the generalised model.  In all cases we consider $z=0.99$.  All squark and slepton masses in the tables below are from the IR-direct gauge mediation contribution; we will discuss the UV mass contributions in the next section.

\subsubsection*{ Spectrum at small Yukawa } 

We consider a case where the Yukawas of (\ref{eq:generalW}) and (\ref{Myukawas}) remain perturbative up to the Landau pole of $SU(3)$; so we consider here the case of $y=2$, $\lambda=1$ and all other Yukawas are $\mathcal{O}(1)$.  As discussed below (\ref{IRgauginoscalarratio}), this results in too light a gluino mass and this situation does not allow for large squark mixing.  We will assume that the UV contributions to the scalar masses have small coefficients so that the flavor diagonal, IR contributions (\ref{IRscalarmass}) dominate.  Solving the RGEs we find the spectrum, at the messenger scale, shown in Table~\ref{tab:smallyuk}: the Landau pole occurs at $\Lambda_3\sim 8\times 10^3~\tev$ and $\alpha_3(M_{mess})\sim 0.12$. 
\begin{table}[h]
\centerline{
\begin{tabular}{c|cc||cc}
$SU(3)$ & $m_{\tilde{q}}$ & $1400~\gev$ & $m_{\tilde{g}}$ & $880~\gev$\\ \hline\hline
$SU(2)$ & $m_{\tilde{l}}$ & $360~\gev$& $m_{\tilde{W}}$ & $520~\gev$\\ \hline\hline
$U(1)$ & $m_{\tilde{e^c}}$ & $160~\gev$ & $m_{\tilde{B}}$ & $370~\gev$\\ \hline\hline
Messenger& $M,M^\prime, \tilde{\Phi}$ & $15~\tev$ & $m_{-}$ &  $10~\tev$\\
sector & $\Mmes$ & $100~\tev$ & $m_\xi$ & $3100~\gev$
\end{tabular}
}
\caption{Spectrum for $y=2$, $\lambda=1$ and all other Yukawas are $\mathcal{O}(1)$. Here and in Tables \ref{tab:largeyuk},\ref{tab:general}, only the IR contributions to squark and slepton masses are shown.}\label{tab:smallyuk}
\end{table}

\subsubsection*{ Spectrum at large Yukawa  } 

As discussed in Section~\ref{irsoft}, to get large gaugino masses and so allow  large sflavor violation in the MRSSM~\cite{Kribs:2007ac} we need a large Yukawa; here we consider  the case of $y=8$ and all other Yukawas are $\mathcal{O}(1)$.  For such a large Yukawa the Yukawa Landau pole is close to the messenger scale.  The squark masses are somewhat large, but below we will assume some cancellation between the UV~(\ref{softscalaroperator}) and IR~(\ref{IRscalarmass}) contributions, allowing for large squark mixing \`a la \cite{Kribs:2007ac}: this will require some tuning and we will discuss this in the next section. In this case, we find the spectrum of Table~\ref{tab:largeyuk} while $\alpha_3(M_{mess})\sim 0.11$ and $\Lambda_3\sim  10^4~\tev$.
The Landau poles of the other SM gauge groups are significantly higher, but as we mentioned above, ``dualizing" color at $\Lambda_{\rm 3}$ would necessarily change that estimate. 
\begin{table}[h]
\centerline{
\begin{tabular}{c|cc||cc}
$SU(3)$ & $m_{\tilde{q}}$ & $1300~\gev$ & $m_{\tilde{g}}$ & $3500~\gev$\\ \hline\hline
$SU(2)$ & $m_{\tilde{l}}$ & $350~\gev$& $m_{\tilde{W}}$ & $2100~\gev$\\ \hline\hline
$U(1)$ & $m_{\tilde{e^c}}$ & $160~\gev$ & $m_{\tilde{B}}$ & $1500~\gev$\\ \hline\hline
Messenger& $M,M^\prime, \tilde{\Phi}$ & $13~\tev$ & $m_{-}$ &  $10~\tev$\\
sector & $\Mmes$ & $100~\tev$ & $m_\xi$ & $13~\tev$
\end{tabular}
}
\caption{Spectrum for $y=8$ and all other Yukawas are $\mathcal{O}(1)$.}\label{tab:largeyuk}
\end{table}
We emphasize that we do not expect this spectrum to be an accurate sample of parameter space with such a large Yukawa coupling; rather, we can see from this exercise that the only hope we have to realize an MRSSM scenario in the IR masses is to go to strong coupling, which would necessitate a more detailed analysis, including the effects of higher loops.

\subsubsection*{Spectrum in the generalized model}

In the models of generalized $R$-symmetric gauge mediation of (\ref{Wtoy}) increasing the number of messenger families, $N_{\rm mess}$, increases the ratio of the gaugino mass to the scalar mass.   Furthermore, the SM Landau pole is postponed because of the absence of the $SU(5)_V$ adjoints $M$, $M^\prime$, which allows us to take a lower messenger scale. Performing the same analysis as above, we find that for $y = 3$, $N_{\rm mess} = 6$ and $M_{\rm mess}=80~\tev$ we have $\alpha_s(M_{\rm mess})= 0.08$ and $\Lambda_3 =  5\times 10^4$ TeV.  The corresponding spectrum is shown in Table~\ref{tab:general}.  Because of the large number of messengers the Yukawa has a Landau pole below $\Lambda_3$.
\begin{table}[h]
\centerline{
\begin{tabular}{c|cc||cc}
 $SU(3)$ & $m_{\tilde{q}}$ & $1900~\gev$ & $m_{\tilde{g}}$ & $5300~\gev$\\ \hline\hline
$SU(2)$ & $m_{\tilde{l}}$ & $620~\gev$& $m_{\tilde{W}}$ & $3500~\gev$\\ \hline\hline
$U(1)$ & $m_{\tilde{e^c}}$ & $290~\gev$ & $m_{\tilde{B}}$ & $2600~\gev$\\ \hline\hline
Messenger sector& $\Mmes$ & $80~\tev$ & & 
\end{tabular}
}
\caption{Spectrum in the generalized model for $y=3$ and $N_{\rm mess}=6$.}\label{tab:general}
\end{table}

\subsection{Estimation of tuning}
\label{tuning}

Recall that there are two contributions to the soft squark masses: one from the direct mediation, which is fixed by the calculation in Section \ref{irsoft}, and the other from the UV operators in (\ref{softscalaroperator}).  The latter comes with a coefficient that we will call $c_D$ for the flavor-diagonal terms, and $c_{OD}$ for the flavor-off-diagonal terms.  Ideally we would like these coefficients to be $\mathcal{O}(1)$, and to solve the flavor puzzle we would also want $c_D\sim c_{OD}$.  This means that there are two potential sources of tuning: one coming from the UV-IR cancellation of the diagonal masses, and one coming from the smallness of the flavor-violating terms relative to the flavor-diagonal terms.  We will discuss each of these in turn.

First of all, some general comments can be made about the first kind of tuning between UV and IR contributions.  Recall that we made the conservative assumption that the scale of the UV operators was proportional to the QCD Landau pole $\Lambda_3$ i.e. $\Lambda=\Lambda_3/4\pi$.  This means that $M_{UV}\sim\Mmes^2/\lambda\Lambda$ is typically quite large unless we wish to make $\lambda$ big, which would introduce another Landau pole.  This mass scale is typically $\mathcal{O}(10)$ TeV in the ISS models, and smaller for the generalized models, as can be seen from the tables in the previous section.  
If the final scalar mass is $m_0$, we have
\be
c_D=\frac{m_0^2-m_{IR}^2}{M_{UV}^2} \label{cD}~,
\ee
with $m_{IR}^2$ given by (\ref{IRscalarmass}).  If $m_0<m_{IR}\sim 1$ TeV, this means that $|c_D|\sim 10^{-2}$ in the ISS models, and $|c_D|\sim 1$ in the generalized models.  This is smaller than hoped for in the ISS case, although it does very well in the generalized model; but it should be noted that it depended on the cutoff being so low, and our hopes to avoid another Landau pole in $\lambda$.  If we are willing to accept strong coupling, or the added assumption that the generation of flavor-changing operators is postponed to a higher scale (the $SU(2)$ Landau pole, for instance), then this tuning can be weakened.

To analyze the second form of tuning, if $\delta$ is the ratio of the flavor-changing mass squared term over $m_0^2$, we have
\be
c_{OD}=\delta\left(\frac{m_0}{M_{UV}}\right)^2 \label{cOD}~.
\ee
Given Equations (\ref{cD})-(\ref{cOD}), we can immediately write down a formula that quantifies the flavor tuning:
\be
t\equiv\left|\frac{c_{OD}}{c_D}\right|=\frac{\delta}{|1-(m_{IR}/m_0)^2|}\label{tuningeq}~.
\ee
Notice that this expression is independent of $M_{UV}$.  Typical allowed values of $\delta$ are of order 0.1 or less \cite{Blechman:2008gu}, given $m_{1/2}/m_0$ of 5--10.  We saw from (\ref{IRgauginoscalarratio}) that $m_{IR}$ is typically larger or of order the gaugino mass, so we immediately see from (\ref{tuningeq}) that this model will be somewhat tuned.  For example, if we demand a 10\% tuning, we require $m_0=m_{IR}/\sqrt{2}$, which is very hard to do while maintaining the gaugino:squark ratio.  Lowering our standard to a 1\% tuning, we require $m_0=m_{IR}/\sqrt{11}$ which is much easier to accomplish.  So there is a trade off.
\begin{table}[h]
\centerline{
\begin{tabular}{c|cc||c}
$ $ & $m_0$ & $\delta$ & $t$ \\
\hline
ISS with Large $y$ & 600 GeV & 0.05 & 1.4\%  \\
General Model & 1 TeV & 0.07 & 2.7\%
\end{tabular}
}
\caption{Size of the flavor tuning for the MRSSM spectra considered above.}\label{tab:tuning}
\end{table}
In Table \ref{tab:tuning} we give the flavor tunings for the two models considered in Tables \ref{tab:largeyuk} and \ref{tab:general}.  The values of $\delta\equiv\delta_L=\delta_R$ are the maximum values for the given $m_0$ and gluino mass after QCD corrections to $K-\bar{K}$ mixing are taken into account \cite{Blechman:2008gu}.

\subsection{Lifetime of the false vacuum}

We have concentrated our attention on the physics around the SUSY breaking vacuum of ISS but this minimum of the potential is metastable.  The true minimum of the system, whose existence is due to the higher dimension non-perturbatively generated term we ignored in (\ref{wmagnetic}), has unbroken supersymmetry.  The additional operator is due to instanton contributions,
\be
W_{inst}=\frac{{\rm det}\CM}{\Lambda^3}~,
\ee
where in this section $\Lambda$ denotes the duality scale, the strong coupling scale of the gauge coupling in the microscopic theory.  Once this additional term is included the rank condition can now be satisfied and there is a SUSY preserving minimum at,
\be
\langle M\rangle\sim f\left(\frac{\Lambda}{f}\right)^{3/5}\unit\, ,\ \langle q\rangle =\langle \bar{q}\rangle =0~.
\ee
Because the additional term is irrelevant this SUSY preserving minimum is far from the SUSY breaking minimum.  It is this distance that results in the metastable vacuum being very long lived.  Transitions from one vacuum to another are initiated by bubble formation, the rate for this process is determined by the action of the 4 dimensional Euclidean bounce action,
\be
\Gamma \sim f^4 \exp\left( -S_4 \right)~.
\ee
In general calculating the bounce action analytically is not possible and it must be determined numerically.  For the case of ISS however the potential is well approximated by a square potential for which there are known analytic solutions~\cite{Duncan:1992ai}.  The bubble action for our model is given~\cite{Intriligator:2006dd,Csaki:2006wi} by
\be
S_4\sim  \left(\frac{\Lambda}{f}\right)^{12/5}~.
\ee
Requiring that the false vacuum lives longer than the age of the universe results in the requirement~\cite{Craig:2006kx}
\be
\left(\frac{\Lambda}{f}\right)\gtap 3~.
\label{lifetime}
\ee
As seen in Section~\ref{lambda} the $SU(3)$ Landau pole, the upper bound on the duality scale, was approximately $100 f$, so (\ref{lifetime}) can be easily satisfied for the scales discussed earlier.

\section{Discussion}

In conventional models of supersymmetry breaking the dynamics that leads to the breaking of supersymmetry also breaks $R$-symmetry.  When this breaking is communicated to the visible sector it results in $R$-symmetry violating gaugino masses, $B_\mu$ and $A$-terms.  There has been much recent interest in the ISS models of supersymmetry breaking for which there exists a metastable supersymmetry breaking vacuum that preserves the $R$-symmetry.  If such models are to be phenomenologically viable the gauginos must acquire a mass.  Many variants of ISS have been explored that break the $R$-symmetry and allow for Majorana gaugino masses.  Here we have discussed the alternative possibility that the $R$-symmetry is preserved and instead the gauginos acquire a Dirac mass.  The Dirac gaugino mass and the sfermion masses are communicated to the visible sector through gauge mediation; hence we have a model of $R$-symmetric Gauge Mediated Supersymmetry Breaking (RGMSB).  Because the $R$-symmetry is preserved the gauginos are Dirac, the $A$-terms are zero, and the Higgs sector now consists of four Higgs doublets: the field content of the MRSSM.  We showed that the dependence of the gaugino mass on the supersymmetry breaking scale differs from that of usual GMSB, but the scalar masses do not.  The gaugino mass is lower than in usual gauge mediation.  

We considered two examples for the $R$-preserving-supersymmetry-breaking sector: a version of ISS which may allow for direct mediation, and a generalization (an O'Raifeartaigh model) with fewer fields.  The necessity of including an adjoint chiral superfield to act as the Dirac partner of the gauginos means that these models have a Landau pole for gauge couplings, the lowest of which is for $SU(3)$.  In the case of the ISS model there are many new fields charged under the standard model and this Landau pole is low, typically a few decades above the scale of the messenger masses.  For the O'Raifeartaigh model it can be somewhat higher.  There are potentially new operators, such as flavor non-diagonal scalar masses, generated at the strong coupling scale.  The size of these operators is unknown.  If small then the model is an $R$-symmetric version of gauge mediation, with a spectrum that differs somewhat from that of \cite{Martin:1996zb}.  However, if large (but not too large) this has all the features of the MRSSM.

Making a conservative estimate of the the size of these UV generated operators we found that it is possible to realize the MRSSM scenario of large flavor-violating couplings by using $R$-symmetric gauge mediation, but only at the expense of introducing fine tuning or strong coupling or both.  If these operators were instead smaller than expected, then the spectrum of the MRSSM could be realized, but there would be no source of the large sfermion mixings (allowed because of the R-symmetry) that lead to the interesting flavor signatures.  This does not rule out the possibility of the MRSSM, but it does suggest that a better understanding of the UV theory is required in order to decide how natural such a spectrum actually is. 

\bigskip
\noindent
{\bf \large Acknowledgements}

 \bigskip

PF and EP thank the KITP Santa Barbara and the Aspen Center for Physics for hospitality during completion of this work. We thank Graham Kribs, Markus Luty, Tim Tait, Martin Schmaltz, and Yuri Shirman for discussions, and Martin Schmaltz for suggesting a title. SA acknowledges the support of the Direcci\'{o}n General de Relaciones
Internacionales de la Secretar\'{i}a de Educaci\'{o}n P\'{u}blica (DGRI-SEP) of Mexico.  
SA, AB, and EP acknowledge support of the National Science and Engineering Research Council (NSERC) of Canada.  Fermilab is operated by Fermi Research Alliance, LLC, under Contract DE-AC02-07CH11359 with the United States 
Department of Energy. 

\appendix
\section{Mass of the pseudo-NG boson} 
\label{NGmass}

As shown in Equations (\ref{psi}) and (\ref{psibar}) there is still a messenger mode that corresponds to the NG boson of the spontaneously broken $U(1)$ symmetry.  However, because of the presence of the operators in $W_1$ this symmetry is explicitly broken\footnote{Notice that if these operators come from the dimension four superpotential term $\bar q  [\hat{\Phi}, \CM] q$ as mentioned in a previous footnote, then these operators no longer explicitly break the symmetry, and in fact the $\xi$ field remains a true massless NG boson.}, and a mass for the $\xi$ mode will be generated at two loops.  To leading order there is only one diagram that generates this mass, shown in Figure \ref{pNGB} plus its complex conjugate.  This diagram is finite.  Expanding around the minimum, with vacuum expectation values from Equations (\ref{psi}) and (\ref{psibar}) and $\langle\eta\rangle=0$, we find the effective potential for $\xi$ is
\be
V_{\rm eff}(\xi)=-\mu^2v^2\cos\left(\frac{2\xi}{v}\right)\nn~,
\ee
where $\mu^2$ is the value of the loop in Figure \ref{pNGB}:
\be
\mu^2=\frac{(\lambda\kappa y)^2}{4}\left[I(m_+,m_+)+I(m_-,m_-)-2I(m_+,m_-)\right]~,
\label{mutwoloop}
\ee
and the Euclidean loop integrals $I(m_1,m_2)$ are computed in \cite{Martin:1996zb} and have the form
\[\int\frac{d^dk}{(2\pi)^d}\int\frac{d^dq}{(2\pi)^d}\left[\frac{1}{(k+q)^2}\frac{1}{k^2+m_1^2}\frac{1}{q^2+m_2^2}\right]~.\]
This leads to a $\xi$ mass:
\be
m_\xi^2=\left(\frac{\lambda\kappa y}{16\pi^2}\right)^2\Mmes^2 H(z)~,
\ee
where
\be
H(z)=(1+z)\left[-\log^2\left(\frac{1+z}{1-z}\right)-2\dilog\left(\frac{-2z}{1-z}\right)\right] + (z\rightarrow -z)~.
\label{hz}
\ee
In particular: $H(1)= 2\pi^2 /3$, and vanishes for $z=0$ (the SUSY limit).

  \bibliographystyle{JHEP}

 \bibliography{paper}

\end{document}